\documentstyle[12pt]{article}

\newcommand{\pl}{\partial} 
\newcommand{\be}{\begin{equation}} 
\newcommand{\ee}{\end{equation}} 
\newcommand{\bea}{\begin{eqnarray}} 
\newcommand{\eea}{\end{eqnarray}} 
\newcommand{\nn}{\nonumber} 
\newcommand{\p}[1]{(\ref{#1})} 

\begin{document}
\setcounter{page}0
\thispagestyle{empty}
\begin{flushright}
hep-th/9602079\\
February 1996
\end{flushright}
\centerline{\bf HARMONIC SUPERSPACE: SOME NEW TRENDS}
\vspace{1truecm}
\vskip0.5cm
\centerline{{\large Evgeny A. Ivanov}}
\vskip.5cm
\centerline{\it Bogoliubov Laboratory of Theoretical Physics, JINR,}
\centerline{\it 141 980, Dubna, Moscow Region, Russian Federation}
\vskip1cm
\begin{center}
{\bf Plenary Talk at the 1st Joint JINR-ROC (Taiwan) Workshop} \\
{\bf ``Intermediate and High Energy Physics''}, \\
{\bf June 26-28, 1995, Dubna, Russia}
\end{center}
\vskip1.5cm
\centerline{\bf Abstract}

\vspace{0.3cm}
\noindent We give a brief account of two recent applications 
of the harmonic superspace method: (i) an off-shell description 
of torsionful $(4,4)$ supersymmetric $2D$ sigma models in the 
framework of $SU(2)\times SU(2)$ harmonic superspace and (ii) 
the harmonic superspace formulation of ``small'' $N=4$ $SU(2)$ 
superconformal algebra and the related super KdV hierarchy.
\vskip.5cm

\newpage
\noindent{\bf 1. Introduction.}
Harmonic superspace has been proposed in 1984 in Dubna as an efficient 
tool for treating theories with extended supersymmetry \cite{GIKOS}. 
This concept allowed to solve the long-standing problem of constructing  
off-shell superfield formulations of all $N=2$, $4D$ supersymmetric 
theories: 
theory of $N=2$ matter (sigma models), $N=2$ super Yang-Mills and 
supergravity theories [1 - 4] as well as of 
$N=3$ super Yang-Mills 
theory \cite{GIKOSn3}. Later on, the same method was applied to purely 
bosonic manifolds to get a new formulation of the Ward construction 
for self-dual Yang-Mills fields \cite{GIOSap1} and to find unconstrained 
potentials for hyper-K\"ahler and quaternionic geometries 
\cite{{GIOSap},{quat}}. 

The essence of the harmonic (super)space approach consists in 
extending the original (super)manifold by some extra variables which 
parametrize the automorphism group of the given (super)manifold. 
The basic advantage of considering 
such extended manifolds is the possibility to single out in them 
a submanifold of lower dimension, the so-called ``analytic subspace''. 
In most examples the 
unconstrained functions on this subspace, analytic (super)fields, turn out 
to be the fundamental entities of the given theory. In particular, 
the potentials underlying the hyper-K\"ahler and quaternionic geometries 
live as unconstrained objects on the relevant analytic subspaces.

Since its invention, the harmonic superspace approach has been further 
advanced and developed along several directions. In the present report we 
review two of its recent developments. 

Firstly, we sketch an off-shell description of torsionful 
$(4,4)$ supersymmetric sigma models in the framework of the 
$SU(2)\times SU(2)$ harmonic superspace \cite{{IS},{Iv},{Iv2}}. 

Secondly, we present the harmonic superspace formulation of ``small'' 
$N=4$ superconformal algebra and the associated $N=4$ KdV hierarchy 
\cite{{DI},{DIK}}. 

\vspace{0.2cm}  
\noindent{\bf 2. SU(2)xSU(2) harmonic superspace.}
The $SU(2)\times SU(2)$ harmonic superspace 
is an extension of the 
standard real $(4,4)$ $2D$ superspace by two independent sets 
of harmonic variables  $u^{\pm 1\;i}$ and $v^{\pm 1\;a}$ 
($u^{1\;i}u^{-1}_{i} =
v^{1\;a}v^{-1}_{a} = 1$) associated with 
the automorphism groups $SU(2)_L$ and $SU(2)_R$ of the left and 
right sectors of $(4,4)$ supersymmetry \cite{IS} (see also \cite{BLR}). 
The most relevant 
{\it analytic} harmonic subspace is spanned by the following set of 
coordinates \be  \label{anal2harm}
(\zeta, u,v) = 
(\;x^{++}, x^{--}, \theta^{1,0\;\underline{i}}, 
\theta^{0,1\;\underline{a}}, u^{\pm1\;i}, v^{\pm1\;a}\;)\;,  
\ee
where we omitted the light-cone indices of odd coordinates. 
The superscript ``$n,m$'' stands for two independent harmonic $U(1)$ 
charges, left ($n$) and right ($m$) ones. The additional doublet indices 
of odd coordinates, $\underline{i}$ and $\underline{a}$, refer to 
two extra automorphism groups $SU(2)_L^{'}$ and $SU(2)_R^{'}$ which 
together with $SU(2)_L$ and $SU(2)_R$ form the full $(4,4)$ 
supersymmetry automorphism group $SO(4)_L \times SO(4)_R$. 

The usefulness of $SU(2)\times SU(2)$ harmonic superspace consists in 
that it allows for a natural off-shell superfield description for 
the important class of $(4,4)$ supersymmetric sigma models, 
those with torsion on the bosonic manifold. This type of sigma models 
is interesting mainly because they can provide non-trivial backgrounds 
for superstrings (see, e.g., \cite{Luest}). 

We start with an off-shell formulation of the twisted $(4,4)$ multiplet 
which until now was the basic object in constructing 
actions for such sigma models (actually, a subclass of them with 
mutually commuting left and right quaternionic structures 
\cite{{GHR},{RSS}}). 
It is described by a real analytic superfield $q^{1,1}(\zeta,u,v)$ 
subjected to the harmonic constraints 
\cite{IS}
\be  \label{qconstr}
D^{2,0} q^{1,1} = D^{0,2} q^{1,1} = 0\;.
\ee
where 
\bea
D^{2,0} &=& \partial^{2,0} + i\theta^{1,0\;\underline{i}}
\theta^{1,0}_{\underline{i}}\partial_{++}\;, 
\;\;D^{0,2} \;=\;\partial^{0,2} + i\theta^{0,1\;\underline{a}}
\theta^{0,1}_{\underline{a}}\partial_{--}  
\nonumber  \\
(\partial^{2,0} &=& u^{1 \;i}\frac{\partial} {\partial u^{-1 \;i}}\;,\;\; 
\partial^{0,2} \;=\; v^{1 \;a}\frac{\partial} {\partial v^{-1 \;a}}) 
\label{harm2der}
\eea
are the left and right mutually commuting analyticity-preserving 
harmonic derivatives. These constraints leave in $q^{1,1}$ $8+8$ 
independent components, that is just the irreducible off-shell 
component content of $(4,4)$ twisted multiplet \cite{{IK},{GHR}}. 

The general off-shell action of $n$ superfields $q^{1,1\;M}$  
$(M=1,2,... n)$ can be written as the following integral over the analytic 
superspace (\ref{anal2harm}) \cite{IS} 
\be
S_{q} = \int \mu^{-2,-2}\; h^{2,2}(q^{1,1\;M},u^{\pm1},v^{\pm1})\;,  
\label{qact}
\ee
$\mu^{-2,-2}$ being the relevant integration measure. 
The analytic superfield lagrangian $h^{2,2}$ is an arbitrary function 
of its arguments (the only restriction on its dependence on the 
harmonics $u$ and $v$ is the consistency with the external 
$U(1)$ charges $2,2$).  

As a non-trivial example of the $q^{1,1}$ action with 
four-dimensional bosonic manifold we quote  
the action of $(4,4)$ extension of the $SU(2)\times U(1)$ 
WZNW sigma model  
\be \label{wzwaction} 
S_{wzw} = \frac{1}{4\kappa ^2} \int \mu^{-2,-2} \;\hat{q}^{1,1} 
\hat{q}^{(1,1)} 
\left(\frac{\mbox{ln}(1+X)}{X^2} - \frac{1}{(1+X)X} \right)\;. 
\label{confact} 
\ee 
Here  
\be 
\hat{q}^{1,1} = q^{1,1} - c^{1,1}\;,\;X = c^{-1,-1}\hat{q}^{1,1}\;,\; 
c^{\pm 1,\pm 1} = c^{ia}u^{\pm1}_iv^{\pm1}_a \;,\; 
c^{ia}c_{ia} = 2\;. 
\ee
Despite the presence of an extra quartet constant $c^{ia}$ in the 
analytic superfield lagrangian, the action (\ref{wzwaction}) 
actually does not depend 
on $c^{ia}$ as it is invariant under arbitrary rigid
rescalings and $SU(2)\times SU(2)$ rotations of this constant. 

The $SU(2)\times SU(2)$ harmonic superspace 
description of 
$(4,4)$ twisted multiplet suggests a new off-shell formulation  
of the latter via unconstrained analytic superfields. After implementing  
the constraints \p{qconstr} in the action with superfield lagrange 
multipliers we arrive at the following new action \cite{IS}
\be
S_{q,\omega} = \int \mu^{-2,-2} \{ 
q^{1,1\;M}(\;D^{2,0} \omega^{-1,1\;M}   + 
D^{0,2}\omega^{1,-1\;M} \;) + h^{2,2} (q^{1,1}, u, v) \}\;.
\label{dualq}
\ee
In (\ref{dualq}) all the involved superfields are unconstrained analytic, 
so from the beginning the action (\ref{dualq}) contains an infinite number 
of auxiliary fields coming from the double harmonic expansions with 
respect to the harmonics $u^{\pm1\;i}, v^{\pm1\;a}$. Varying 
with respect to the Lagrange multipliers $\omega^{1,-1\;M}, 
\omega^{-1,1\;M}$ takes one back to the action \p{qact} and 
constraints \p{qconstr}. On the other hand, varying with respect to 
$q^{1,1\;M}$ yields an algebraic equation for eliminating this superfield, 
which enables one to get a new dual off-shell representation of the twisted 
multiplet action through unconstrained analytic superfields 
$\omega^{-1,1\;M}$, $\omega^{1,-1\;M}$. 

The crucial feature of the action (\ref{dualq}) (and its $\omega$ 
representation) is the abelian gauge invariance 
\be  
\delta \;\omega^{1,-1\;M} = D^{2,0} \sigma^{-1,-1\;M} 
\;, \; \delta \;\omega^{-1,1\;M}  
= - D^{0,2} \sigma^{-1,-1\;M}\;,
\label{gauge}
\ee
where $\sigma^{-1,-1\;M}$ are unconstrained analytic 
superfield parameters. This gauge freedom ensures the 
on-shell equivalence of the $q,\omega$ or $\omega$ formulations of 
the twisted multiplet action to its original $q$ formulation 
\p{qact}: it 
neutralizes superfluous physical dimension component fields in the 
superfields $\omega^{1,-1\;M}$ and $\omega^{-1,1\;M}$ and 
thus equalizes the numbers of propagating fields in both formulations. It 
holds already at the free level, with 
$h^{2,2}$ quadratic in $q^{1,1\;M}$, so it is natural to expect that 
any reasonable generalization of the action (\ref{dualq}) respects 
this symmetry or a generalization of it.  
  
It is well known that with making use of $(4,4)$ twisted multiplets 
one may construct invariant off-shell actions only for those 
torsionful $(4,4)$ 
sigma models in which left and right triplets of the covariantly constant 
complex structures on the bosonic target mutually commute 
\cite{{GHR},{RSS}}. It turns out that the above dual form of the 
twisted multiplet action is 
a good starting point for constructing more general actions which 
admit no inverse duality transformation to the twisted multiplets ones
and provide an off-shell description of sigma models with non-commuting 
left and right complex structures. 

These more general actions are obtained by allowing for the dependence 
on the superfields $\omega$s also in $h^{2,2}$. With making use of 
freedom of reparametrizations in the target space together with the 
self-consistency constraints following from the commutativity condition  
\be \label{comm} 
[\;D^{2,0}, D^{0,2}\;] = 0\;,
\ee
one may show that the most general action of the analytic superfields 
triples $\omega^{1,-1\;M}$, $\omega^{-1,1\;M}$, $q^{1,1\;M}$ can be 
reduced to the form \cite{{Iv},{Iv2}}
\bea
S_{q,\omega} &=& 
\int \mu^{-2,-2} \{\; q^{1,1\;M}D^{0,2}\omega^{1,-1\;M} + 
q^{1,1\;M}D^{2,0}\omega^{-1,1\;M} +  \omega^{1,-1\;M}h^{1,3\;M} 
\nonumber \\
&&+ \omega^{-1,1\;M}h^{3,1\;M} + \omega^{-1,1\;M} \omega^{1,-1\;N}
\;h^{2,2\;[M,N]} + h^{2,2}\;\}\;, \label{haction}
\eea
where the involved potentials depend only on $q^{1,1\;M}$ and target 
harmonics and satisfy the following constraints  
\bea 
&& \nabla^{2,0} h^{1,3\;N} - \nabla^{0,2} h^{3,1\;N} + h^{2,2\;[N,M]} 
\;\frac{\partial h^{2,2}}{\partial q^{1,1\;M}} \;=\; 0 \label{1} \\
&& \nabla^{2,0} h^{2,2\;[N,M]} - 
\frac{\partial h^{3,1\;N}}{\partial q^{1,1\;T}} \;h^{2,2\;[T,M]} + 
\frac{\partial h^{3,1\;M}}{\partial q^{1,1\;T}}\; h^{2,2\;[T,N]} \;=\; 0 
\label{2} \\
&& \nabla^{0,2} h^{2,2\;[N,M]} - 
\frac{\partial h^{1,3\;N}}{\partial q^{1,1\;T}}\; h^{2,2\;[T,M]} + 
\frac{\partial h^{1,3\;M}}{\partial q^{1,1\;T}}\; h^{2,2\;[T,N]} \;=\; 0 
\label{3} \\
&& h^{2,2\;[N,T]}\;\frac{\partial h^{2,2\;[M,L]}}{\partial q^{1,1\;T}} + 
h^{2,2\;[L,T]}\;\frac{\partial h^{2,2\;[N,M]}}{\partial q^{1,1\;T}} +
h^{2,2\;[M,T]}\;\frac{\partial h^{2,2\;[L,N]}}{\partial q^{1,1\;T}} 
\;=\; 0\;, 
\label{4} 
\eea
\be
\nabla^{2,0} = \partial^{2,0} + h^{3,1\;N}\frac{\partial}{\partial 
q^{1,1\;N}}
\;,\;\; 
\nabla^{0,2} = \partial^{0,2} + h^{1,3\;N}\frac{\partial}{\partial 
q^{1,1\;N}}
\;.
\ee
Here $\partial^{2,0}, \partial^{0,2}$ act only on the target harmonics.

The action (\ref{haction}), with taking account of the 
constraints (\ref{1}) - (\ref{4}), enjoys the following non-abelian 
generalization of the gauge freedom \p{gauge} 
\bea
\delta \omega^{1,-1\;M}  &=&
\left( D^{2,0}\delta^{MN} + 
\frac{\partial h^{3,1\;N}}{\partial q^{1,1\;M}}
\right) \sigma^{-1,-1\;N} - \omega^{1,-1\;L} \;
\frac{\partial h^{2,2\;[L,N]}}
{\partial q^{1,1\;M}}\;\sigma^{-1,-1\;N}\;, 
\nonumber \\
\delta \omega^{-1,1\;M}  &=&
- \left( D^{0,2}\delta^{MN} + \frac{\partial h^{1,3\;N}}{\partial 
q^{1,1\;M}} \right) \sigma^{-1,-1\;N} - \omega^{-1,1\;L} \;
\frac{\partial h^{2,2\;[L,N]}}{\partial q^{1,1\;M}}\;
\sigma^{-1,-1\;N}\;, \nonumber \\
\delta q^{1,1\;M} &=& \sigma^{-1,-1\;N} h^{2,2\;[N,M]}\;. 
\label{gaugenab}
\eea
In general, these gauge transformations close with a field-dependent 
Lie bracket parameter: 
\be
\delta_{br} q^{1,1\;M} = \sigma^{-1,-1\;N}_{br} h^{2,2\;[N,M]}\;, \;\;
\sigma^{-1,-1\;N}_{br} = -\sigma^{-1,-1\;L}_1 \sigma^{-1,-1\;T}_2 
\frac{\partial h^{2,2\;[L,T]}}{\partial q^{1,1\;N}}\;.
\ee
We see that eq. (\ref{4}) guarantees the nonlinear closure of the 
algebra of gauge transformations (\ref{gaugenab}) and so it is a group 
condition similar to the Jacobi identity. 
Remarkably, these gauge transformations augmented with the group 
condition (\ref{4}) 
are precise bi-harmonic counterparts of the two-dimensional version of 
basic relations of 
the so-called Poisson nonlinear gauge theory 
which recently received some attention 
\cite{Ikeda}. The manifold $(q,u,v)$ can be 
interpreted as a kind of bi-harmonic extension of some Poisson 
manifold and the potential 
$h^{2,2\;[N,M]}(q,u,v)$ as a tensor field inducing the Poisson 
structure on this extension. 

It should be pointed out that it is the presence of the 
antisymmetric potential $h^{2,2\;[N,M]}$ that makes the considered case 
non-trivial and, in particular, the gauge invariance (\ref{gaugenab}) 
non-abelian. If $h^{2,2\;[N,M]}$ is vanishing, (\ref{haction}) proves  
to be reducible to the dual action of twisted multiplets (\ref{dualq}). 
In the $n=1$ case the potential 
$h^{2,2\;[N,M]}$ vanishes identically, so (\ref{haction}) 
for $n=1$ is actually equivalent to (\ref{dualq}). 
Thus {\it only for} $n\geq 2$ a new 
class of torsionful $(4,4)$ sigma models comes out. 

It is easy to see that the action (\ref{haction}) with non-zero 
$h^{2,2\;[N,M]}$ {\it does not} admit any duality transformation 
to the form with the superfields $q^{1,1\;M}$ only. 
So, the obtained system certainly does not admit in general an 
equivalent description in terms of twisted multiplets and, 
for this reason, the left and right complex structures on the 
target space can be non-commuting. In refs. \cite{{Iv},{Iv2}} 
we have explicitly shown this non-commutativity for a 
particular class of the models in question. 

This subclass corresponds to the choice of the following ansatz
\bea
h^{1,3\;N} &=& h^{3,1\;N} \;=\; 0\;; \; h^{2,2} \;=\; 
h^{2,2}(t,u,v)\;, \; 
\;t^{2,2} \;=\; q^{1,1\;M}q^{1,1\;M}\;; \nonumber \\
h^{2,2\;[N,M]} &=& b^{1,1} f^{NML} q^{1,1\;L}\;, \;b^{1,1} \;=\; 
b^{ia}u^1_iv^1_a\;, \; b^{ia} = \mbox{const}\;,
\label{solut}
\eea
where the real constants $f^{NML}$ are totally antisymmetric. 
The constraints (\ref{1}) - (\ref{3}) are identically satisfied 
with this ansatz, while (\ref{4}) is now none other than the 
Jacobi identity which tells us that the constants 
$f^{NML}$ are structure constants of some real semi-simple Lie 
algebra. Thus the associated $(4,4)$ sigma models can be interpreted 
as a kind of Yang-Mills theories in the harmonic superspace. 
They provide the direct non-abelian generalization 
of the twisted multiplet sigma models with the action (\ref{dualq}) 
which are thus analogs of two-dimensional abelian gauge theory. 
The action (\ref{haction}) specialized to the  
case (\ref{solut}) is as follows
\bea
S^{YM}_{q,\omega} &=& 
\int \mu^{-2,-2} \{\; q^{1,1\;M} (\; D^{0,2}\omega^{1,-1\;M} + 
D^{2,0}\omega^{-1,1\;M} + b^{1,1} \;\omega^{-1,1\;L} \omega^{1,-1\;N}
f^{LNM}\; ) \nonumber \\
&&+ \;h^{2,2}(q,u,v) \} \;.\label{haction0} 
\eea
It is a clear analog of the Yang-Mills action in the first order 
formalism, $q^{1,1\;N}$ being an analog of the YM field strength 
and $b^{1,1}$ of the YM coupling constant. 

An interesting specific feature of this ``harmonic Yang-Mills theory'' 
is that the ``coupling constant'' $b^{1,1} = b^{ia}u^1_iv^1_a$ 
is doubly charged (this is necessary for the balance of harmonic 
$U(1)$ charges). When $b^{ia} \rightarrow 0$, the non-abelian structure 
contracts into the abelian one and we reproduce the twisted multiplet 
action (\ref{dualq}). It also turns out that $b^{ia}$ measures the `
`strength'' of non-commutativity of the left and right complex 
structures on the bosonic target. 

In the simplest case of $h^{2,2} = q^{1,1\;M} q^{1,1\;M}$ we have 
calculated \cite{{Iv},{Iv2}}, in the first non-vanishing order in fields, 
the bosonic metric and torsion potential, as well as the relevant 
triplets of complex structures. All these proved to be represented by 
non-trivial expressions, the commutator of the left and right complex 
structures being non-vanishing and proportional to $b^{ia}$.

Thus in the present case in the bosonic sector we encounter 
a more general geometry compared to the one associated with 
twisted $(4,4)$ multiplets. The basic characteristic 
feature of this geometry is the non-commutativity  of the left and right 
complex structures. It is related in a puzzling way to 
the non-abelian structure of the analytic superspace 
actions (\ref{haction0}), (\ref{haction}): the coupling constant 
$b^{1,1}$ (or the Poisson potential $h^{2,2\;[M,N]}$ in the general case) 
measures the strength of the non-commutativity of complex structures. 

In summary, within the $SU(2)\times SU(2)$ harmonic superspace approach 
it becomes possible to solve the long-standing problem of manifestly 
supersymmetric 
off-shell description of the torsionful $(4,4)$ supersymmetric sigma 
models with non-commuting left and right complex structures.

\vspace{0.2cm}  
\noindent
{\bf 3. Harmonic superspace formulation of N=4 SU(2) super KdV hierarchy.} 
The supercurrent generating ``small'' $N=4$ $SU(2)$ superconformal algebra 
(SCA) admits a concise formulation in the $1D$ $SU(2)$ analytic harmonic 
superspace
\be 
(\zeta, u^{+i}, u^{-j}) = (z, \theta^{+}, \bar \theta^{+}, 
u^{+i}, u^{-j})\;, 
\ee
where it is represented by a doubly charged superfield 
$V^{++}(\zeta, u)$ subjected to the constraint \cite{{a12},{DI}}
\be
D^{++} V^{++} = 0\;. \label{ct3}
\ee
with 
$$
D^{++} = \pl^{++} - i\theta^+ \bar \theta^+ \partial_z\;, \; \pl_{++} 
\equiv u^{+i}\frac{\pl}{\pl u^{-i}}\;.
$$
Solving \p{ct3}, one gets 
\be
V^{++} (\zeta) = w^{ij} u^+_iu^+_j  - \theta^+ \xi^k  u^+_k 
+ \bar \theta^+ \bar \xi^k u^+_k + \theta^+\bar \theta^+ 
\left( i \pl w^{ik} u^+_iu^-_k + T  \right)\;.
\ee
Here, up to scaling factors, the component fields coincide with 
the currents of $N=4\;SU(2)$ SCA: 
the $SU(2)$ triplet of spin 1 currents generating 
$SU(2)$ affine Kac-Moody subalgebra, a complex doublet of fermionic   
spin 3/2 currents and the spin 2 conformal stress-tensor, 
respectively \cite{a9}. 

The standard commutation relations for these currents follow from the 
following superfield Poisson bracket \cite{DI}
\bea  \left\{ V^{++}(1),
 V^{++}(2)\right\} &=& {\cal D}^{(++|++)} \Delta (1 - 2) \nn \\
{\cal D}^{(++|++)} &\equiv & (D^+_1)^2(D^+_2)^2 \left(
\left[ \left({u^+_1u^-_2
\over u^+_1u^+_2}\right) - {1\over 2}D^{--}_2\right]
 V^{++}(2)
- {k\over 4}\pl_{2}\right),
\label{poi}\eea
where $\Delta(1-2)=\delta(x_1-x_2)\;(\theta^1-\theta^2)^4$ is the
ordinary $1D \;N=4$ superspace delta function and 
$$
(D^+)^2 \equiv D^+ \bar D^+ 
$$
are operators ensuring the harmonic analyticity with respect to 
both superspace arguments, $D^+, \bar D^+$ being the relevant 
$u$ projections of $1D$, $N=4$ spinor derivatives. Note that the 
harmonic singularity in the r.h.s. of \p{poi} is fake: it is cancelled 
after decomposing the harmonics $u^{\pm i}_2$ over $u^{\pm i}_1$ with 
making use of their completeness relation. 

To deduce the super KdV equation with the second
hamiltonian structure given by the $N=4\;SU(2)$ SCA in the form \p{poi} 
we need to construct the relevant hamiltonian of the dimension 3.
The most general dimension 3 
$N=4$ supersymmetric hamiltonian $H_3$ one may construct out
of $V^{++}$ reads
\be
H_3 =\int [dZ]\;V^{++}(D^{--})^2V^{++}-i\int [d\zeta^{-2}]\;
c^{-4}(u)\;(V^{++})^3
\;.
\label{h3}
\ee
Here $[dZ]$ and $[d\zeta^{-2}]$ are the integration 
measure of the full harmonic superspace and its analytic subspace, 
$D^{--}$ is the second harmonic derivative (not preserving the 
analyticity) such that 
$$
[\;D^{++}, D^{--}\;] = D^{0}\;, 
$$
$D^{0}$ being the operator counting the harmonic $U(1)$ charge, e.g., 
$D^{0} V^{++} = 2 V^{++}$. The $U(1)$ invariance of the integral over 
analytic subspace requires the inclusion of the harmonic monomial
$c^{-4}(u)=c^{ijkl}u^-_iu^-_ju^-_ku^-_l$ which explicitly breaks $SU(2)$
symmetry. The coefficients $c^{ijkl}$  belong to the dimension 5 spinor
representation of $SU(2)$, i.e. form a symmetric traceless rank 2 tensor, 
and completely break the
$SU(2)$ symmetry, unless $c^{-4}$ is of the
special form 
\be  \label{strc4}
c^{-4}(u)=(a^{-2}(u))^2\;, \;\;\;
a^{-2}(u)=a^{ij}u^-_iu^-_j\;.
\ee 
After taking off the harmonics this condition becomes 
\be
c^{ijkl} = {1\over 3} \left( a^{ij} a^{kl} + a^{ik}a^{jl} + 
a^{il}a^{jk} 
\right)\;.  \label{consa1} 
\ee 
In this case, the symmetry
breaking parameter belongs to the dimension 3 (vector) representation of 
$SU(2)$, and thus has $U(1)$ as a little group. 

Using the hamiltonian \p{h3}, we construct the relevant 
evolution equation:
\be  V^{++}_t = \left\{ H_3, V^{++} \right\}\;.\ee
In the explicit form it reads
\bea
 V^{++}_t&=&i \left( D^+ \right)^2 \left\{ {k\over 2}D^{--}V^{++}_{xx}
-\left[V^{++}(D^{--})^2V^{++}-{1\over 2}(D^{--}V^{++})^2\right]_x 
\right.\cr
&& \left. -{3\over 20}k A^{-4}(V^{++})^2_x+{1\over 2}A^{-6}(V^{++})^3
\right\}\;. \label{kdv4}
\eea
Here $A^{-4}$ and $A^{-6}$ are differential operators on the 2-sphere
$\sim SU(2)/U(1)$
\bea
A^{-4}&=&\sum_{N=1}^4(-1)^{N+1}c^{2N-4}{1\over N!}(D^{--})^N,\cr
A^{-6}&=&{1\over 5}\sum_{N=0}^4(-1)^{N}c^{2N-4}{(5-N)
\over (N+1)!}(D^{--})^{N +1}\;.\eea
We have used the notation:
\be c^{2N-4}={(4-N)!\over 4!}(D^{++})^Nc^{-4}, \ N=0\cdots 4\;.\ee

Equation \p{kdv4} is the $N=4$ $SU(2)$ super KdV equation. 
It is easy to check that its r.h.s is analytic and satisfies 
the same constraint \p{ct3} as the l.h.s. 

As is known, the $N=2$ super KdV equation is integrable 
only for $a=4,\;-2,\;1$, $a$ being a free parameter  in the 
corresponding hamiltonian [20 - 24]. Since the $SU(2)$ breaking tensor 
$c^{ijkl}$ is a direct analog of this $N=2$ KdV parameter 
(and is reduced to it upon the reduction $N=4 \rightarrow N=2$), 
one may expect that the $N=4$ super KdV equation is integrable 
only when certain restrictions are imposed on this tensor. 
To see which kind of restrictions arises, in \cite{{DI},{DIK}} 
we examined the issue of the existence of 
non-trivial conserved charges for \p{kdv4} which are in involution  
with the hamiltonian \p{h3}. Let us sketch the results of this 
analysis.

Conservation of the dimension 1 charge :
\be H_1=\int [d\zeta^{-2}]\; V^{++} \label{h1} \ee
imposes no condition on the parameters of the hamiltonian.

A charge with dimension 2 exists only provided the condition \p{strc4} 
holds. It reads:
\be H_2= i\int [d\zeta^{-2}]\; a^{-2}\;(V^{++})^2 \;.\ee
The conservation of this charge implies a stringent constraint 
on $a^{ij}$, namely 
\be a^{+2}a^{-2}-(a^0)^2={1\over 2}\;a^{ij}a_{ij}=
-{10\over k}\;,
\label{ct4} \ee
where
$$
a^{+2} = D^{++} a^0 = {1\over 2}\;(D^{++})^2 a^{-2} = a^{ij}u^+_iu^+_j\;.
$$
Assuming that the central charge $k$ is integer (if
we restrict ourselves to unitary representations of the $SU(2)$ 
Kac-Moody algebra \cite{wit}), eq. \p{ct4} means that $a^{ij}$ 
parametrizes some sphere $S^{2} \sim SU(2)/U(1)$, such that the 
reciprocal of its radius is {\it quantized}. 

The next conserved charge is a 
dimension 4 one $H_4$ (the dimension 3 conserved charge is the 
$N=4$ KdV hamiltonian itself). It exists under the same restrictions 
\p{strc4}, \p{ct4} on  $c^{ijkl}$ and reads:
\be H_4=\int [dZ]\;a^{-2} V^{++}(D^{--}V^{++})^2
+{i\over 6}\int [d\zeta^{-2}] \left[{7\over 6}\;(a^{-2})^3 (V^{++})^4 -k
a^{-2} ( V^{++}_x)^2\right]. \label{ct5}
\ee
It is curious that it yields the same $N=4$ KdV equation \p{kdv4} via 
the first hamiltonian structure associated with the Poisson bracket 
\be
\left\{V^{++}(1), V^{++}(2)\right\}_{(1)}= i
\left( a^{0}(1) - a^{+2}(1){u^-_1u^+_2\over u^+_1u^+_2}
\right)
(D^+_1)^2 (D^+_2)^2
\Delta(1-2) \label{poi1}\;.
\ee
This bracket is related to 
the original one \p{poi} via the shift
\be V^{++}\longrightarrow V^{++}+ i a^{+2}(u)\;.
\ee
Taking as a new hamiltonian
\be
H_{(1)}= -i {9k\over 4}\;H_4\;,\label{ha1}
\ee
we can reproduce \p{kdv4} as the hamiltonian flow:
\be V^{++}_t=\{H_{(1)},V^{++}\}_{(1)} \;. \label{kdv41} \ee 
This comes about in a very non-trivial way, an essential use of the 
constraint \p{ct4} has to be made in the process.

Thus the conditions \p{strc4} and \p{ct4}  
are necessary not only 
for the existence of the first non-trivial conservation laws for 
eq. \p{kdv4}, but also for it to be bi-hamiltonian. This property 
persists for the evolution equations associated with other 
conserved charges. For instance, the equation associated with $H_2$ 
via the structure \p{poi}, with respect to \p{poi1} has $H_3$ as the 
hamiltonian.

The last bosonic conserved charges we have constructed until now 
\cite{DIK} are $H_5$ and $H_6$. Once again, they exist iff the conditions 
\p{strc4} and \p{ct4} are satisfied. They possess rather complicated 
structure. For instance, $H_5$ is given by the following 
expression (without loss of generality, we have chosen $k=2$)
\bea 
H_5 &=& {1\over 2} \int [dZ] \left[ {1\over 4} \left( D^{--}V^{++} 
\right)^4 + i \left( D^{--}V^{++} \right)^2 
\left( D^- \right)^2 V^{++} 
\right. \nn \\
&& + \; \left. {15\over 4}\; (a^{-2})^2 \left( D^{--}V^{++} \right)^2
\left( V^{++} \right)^2  
- {1\over 2} \left( D^{--}V^{++}_x \right)^2\right]  \nn \\ 
&& + {i\over 4} \int [d\zeta^{-2}] \left[ {63\over 100}\; (a^{-2})^4 
\left( V^{++} \right)^5 - 5 \;(a^{-2})^2 \left( V^{++}_x 
\right)^2 V^{++} 
\right] \;. \label{harmh5}
\eea  

Both the bi-hamiltonian structure and the existence of 
non-trivial 
conserved charges are indications that $N=4$ KdV equation \p{kdv4} 
with the restrictions \p{strc4}, \p{ct4} is integrable, i.e. gives rise 
to a whole $N=4$ $SU(2)$ super KdV hierarchy which is unique. Clearly, 
for the rigorous proof one should, before all, either find the relevant 
Lax pair or show the existence of an infinite number of conserved 
charges of the type given above (e.g., by employing recursion relations 
implied by the bi-hamiltonian property \cite{{Mag},{Opop}}). These 
problems still remain to be solved. 

After reduction $N=4 \rightarrow N=2$ at the superfield 
level \cite{{DI},{DIK}}, $N=4$ $SU(2)$ SCA breaks down to the $N=2$ SCA 
and one is left with the standard $N=2$ KdV, the free parameter $a$ of the 
latter being identified with  
\be
a = {6\over 5} c^{1212} = {2\over 5} \left( a^{11}a^{22} + 2 a^{12}a^{12} 
\right)\;, \; a^{ij}a_{ij} =  - 2 \;(a^{12} a^{12} - a^{11} a^{22}) = 
-10 \;.\label{defa} 
\ee
Actually, only for 
three independent fixings of the $SU(2)$ breaking 
parameter $a^{ik}$ (corresponding to the three independent 
choices of the $SU(2)$ frame), the consistent reduction to 
$N=2$ KdV turns out to be possible \cite{DIK}. For these three choices 
the constraint \p{ct4} gives rise, through the relations \p{defa}, to 
the following values of the $N=2$ KdV parameter
\be \label{valuesab}
(a)\; a= 4; \;\;\; \;(b)\; a = -2; \;\;\; (c)\; a= -2 \;. 
\ee
In other words, we get two $N=2$ super KdV hierarchies, with $a=4$ and 
$a=-2$, as two inequivalent reductions of the single $N=4$ $SU(2)$ 
hierarchy (the cases $(b)$ and $(c)$ yield the same system).

To summarize, the $1D$ $N=4$ harmonic superspace approach allows us to 
construct, in a manifestly supersymmetric way, the super KdV equation 
associated with $N=4$ $SU(2)$ SCA, to reveal the existence of higher-order 
conservation laws and bi-hamiltonian structure for this system, and 
to show that it embodies as two different subsystems two of the three 
inequivalent $N=2$ super KdV hierarchies. 
 
\vspace{0.2cm}
\noindent{\bf 4. Perspectives.}
As regards the new class of $(4,4)$ sigma models described 
in Sect. 2, the obvious problems for further study  
are to compute the relevant metrics and torsions in a closed form 
and to try to utilize the corresponding manifolds as backgrounds for 
some superstrings. An interesting question is 
as to whether the constraints 
(\ref{1}) - (\ref{4}) admit solutions corresponding to the $(4,4)$ 
supersymmetric group 
manifold WZNW sigma models \cite{belg}. 

It still remains to examine whether the 
action (\ref{haction}) 
indeed describes most general $(4,4)$ models with torsion. 
The constrained superfield 
$q^{1,1\;M}$ the dual action of which was a starting point of 
our construction, actually 
represents only one type of $(4,4)$ twisted 
multiplet \cite{IK}. There exist other 
types which differ in the $SU(2)_L\times SU(2)_R$ 
assignment of their component fields 
\cite{{GHR},{IKL},{GI}}. At present it is unclear how 
to simultaneously describe all of them 
in the framework of the same $SU(2)\times SU(2)$ analytic  
harmonic superspace. Perhaps, their actions are related to 
those of $q^{1,1}$ 
by a kind of duality transformation. It may happen, however, 
that for their self-consistent description one will need a 
more general type of $(4,4)$ harmonic superspace, 
with the whole $SO(4)_L\times SO(4)_R$ automorphism group 
of $(4,4)$ supersymmetry harmonized. The related actions will be 
certainly more general than those presented here. 

Among the problems related to the  
$N=4$ super KdV hierarchy, let us mention, besides the 
construction of a Lax pair representation for the $N=4$ $SU(2)$ KdV, a 
generalization to the case of the ``large'' 
$N=4$ SCA with the affine subalgebra $so(4) \times u(1)$ 
\cite{{belg1},{IKL}}. The corresponding $N=4$ super KdV 
hierarchy is expected to embrace both the $N=4$ $SU(2)$ and $N=3$ KdV 
ones as particular cases. Since the large $N=4$ SCA can be viewed as 
a closure of two ``small'' $N=4$ SCAs, each entering with its own affine 
$su(2)$ subalgebra, one may conjecture that a $1D$ version of 
$SU(2)\times SU(2)$ harmonic superspace (or its further extension) 
can be relevant to this problem.

\vspace{0.5cm}
\noindent{\Large\bf Acknowledgements} 

\vspace{0.2cm}
\noindent The author thanks Organizers of the Dubna - Taiwan Workshop 
for the suggestion to give this talk.

\end{document}